\documentclass[cits]{PoS}
\usepackage{amsmath}
\usepackage{fancyvrb}
\usepackage{caption}
\usepackage{subcaption}

\def\la{\langle}
\def\ra{\rangle}

\def\spA#1#2{\la#1#2\ra}
\def\spB#1#2{[#1#2]}
\def\spAB#1#2#3{\la#1|#2|#3]}

\DeclareMathOperator{\tr}{\rm tr}
\def\trm{\tr_-}
\def\trp{\tr_+}

\def\eps{\epsilon}

\def\MP#1#2{#1\cdot #2}
\def\cv#1#2{\spAB{#1}{\gamma^\mu}{#2}}

\title{Multi-loop integrand reduction techniques}

\ShortTitle{Multi-loop integrand reduction}

\author{\speaker{Simon Badger}$^a$, Hjalte Frellesvig$^{b,c}$ and Yang Zhang$^{c}$ \\
  \llap{$^a$}Theory Division, Physics Department, CERN, CH-1211 Geneva 23, Switzerland \\
  \llap{$^b$}Instituto Nazionale di Fisica Nucleare, Sezione di Roma, P.le Aldo Moro 2, 00185 Roma, Italy \\
  \llap{$^c$}Niels Bohr International Academy and Discovery Center, The Niels Bohr Institute,
  University of Copenhagen, Blegdamsvej 17, DK-2100 Copenhagen, Denmark \\
  E-mail: \email{simon.badger@cern.ch}}

\abstract{We review recent progress in $D$-dimensional integrand reduction algorithms
for two loop amplitudes and give examples of their application to non-planar
maximal cuts of the five-point all-plus helicity amplitude in QCD.\\[1cm]
CERN-PH-TH-2014-128
}

\FullConference{Loops and Legs in Quantum Field Theory - LL 2014,\\
		 27 April - 2 May 2014 \\
		 Weimar, Germany }

\begin{document}

\section{Introduction}

Multi-leg two-loop amplitudes are of potential importance for precision
measurements in the coming years of high energy proton-proton collisions at the
LHC. Improved understanding of IR subtraction schemes has led to considerable
progress in the calculations of full NNLO QCD corrections for $2\to2$ processes. Recent examples
of hadronic production of di-jets \cite{Ridder:2013mf,Currie:2013dwa}, $t\bar{t}$ \cite{Abelof:2014fza,Czakon:2013goa}, and $VV$ \cite{Cascioli:2014yka} are reviewed in these
proceedings. The double virtual corrections required at this precision have mainly been obtained using Feynman
diagrams together with integration-by-parts identities \cite{Chetyrkin:1981qh},
but owing to the rapid growth in complexity this approach may not be
sufficient to handle higher multiplicity processes and therefore new techniques are
desirable.

Though Feynman diagram technology has been sufficient for a number of two-loop
QCD amplitudes (those required for di-jets and Higgs plus jet production are
arguably the most complicated amplitudes achieved with this approach
\cite{Anastasiou:2000kg,Anastasiou:2000ue,Anastasiou:2001sv,Glover:2001af,Gehrmann:2011aa}),
on-shell approaches can avoid large intermediate steps and enable an efficient
calculation of more complicated processes. Unitarity~\cite{Bern:1994zx} and
generalized unitarity~\cite{Britto:2004nc} techniques have been successfully
applied to two-loop QCD amplitudes for massless $2\to2$ processes
\cite{Bern:2000dn,Bern:2000ie,Bern:2001df,
Bern:2001dg,Bern:2002tk,Bern:2002zk,Bern:2003ck}. In super-symmetric gauge and
gravity theories these techniques are now a familiar technology, with the
current state-of-the art computations being able to handle four and even five loops.

Encouraged by the high levels of automation achieved at NLO, there has been
recent progress in extending unitarity, generalized unitarity and integrand reduction
algorithms to allow a systematic algebraic approach to arbitrary loop amplitudes. The
maximal unitarity approach proposed by Kosower and Larsen \cite{Kosower:2011ty}
builds upon the direct computation, and further developments in this direction
are summarized in
\cite{Johansson:2012zv,Johansson:2013sda,CaronHuot:2012ab,Larsen:2012sx,Sogaard:2013yga,Sogaard:2013fpa,Sogaard:2014ila,Sogaard:2014oka}.
The integrand reduction algorithm developed by Ossola, Papadopoulos and Pittau
(OPP)~\cite{Ossola:2006us} has also been the focus of multi-loop extensions. Initial attempts to extend this method
\cite{Mastrolia:2011pr,Badger:2012dp} led to the proposal of the computational
algebraic geometry method \cite{Zhang:2012ce, Mastrolia:2012an}, generalizing
the integrand reduction algorithm systematically to all loop orders. A number
of different examples have been considered within this framework
\cite{Badger:2012dv,Mastrolia:2012wf,Kleiss:2012yv,Feng:2012bm,Mastrolia:2013kca,Huang:2013kh,Badger:2013gxa}.
Investigations into spinor integration methods at two-loops are also on-going
\cite{Feng:2014nwa}.

It has been interesting to see algebraic geometry play an increasingly
important role in understanding the details of these methods.
Gr\"obner basis and polynomial division techniques allow the automation of
the integrand reduction process, and the tool of primary decomposition
characterizes the structure of the branches of the unitarity-cut solutions.

In these proceedings we review the $D$-dimensional formulation of the multi-loop integrand reduction
method and present applications to maximal non-planar cuts of the five-gluon all-plus amplitude in QCD.

\section{A $D$-dimensional integrand reduction algorithm}

The details of our approach to $D$-dimensional integrand reduction have been
developed during the computation of the five-gluon all-plus helicity amplitude
in QCD \cite{Badger:2013gxa}. A summary of the approach has recently been
presented in the proceedings of ACAT 2013 \cite{Badger:2013sta}, so we will
only give a very schematic overview here.

A $D=4-2\eps$ dimensional $L$ loop amplitude depending on a set of external momenta $\{p\}$ and internal momenta $\{k\}$ has the generic form:
\begin{align}
  A_n^{(L),[D]}(\{p\}) &= \int \prod_{i=1}^{L} \frac{d^{D} {k_i}}{(2\pi)^D}
  \frac{N(\{k\},\{p\})}{\prod_{l=1}^{L(L+9)/2} D_l(\{k\},\{p\})} \;,
\end{align}
where $D_l$ are the denominators of the loop propagators. The master numerator function $N$ may be obtained from Feynman diagrams or by off-shell recursive
techniques. Alternatively can $N$ be written as the product of tree-level amplitudes
when generalized unitarity cuts are applied to the propagators. The goal of the integrand
reduction procedure is to write a loop amplitude in the form:
\begin{align}
  A_n^{(L),[D]}(\{p\})
   &= \int \prod_{i=1}^{L} \frac{d^{D} {k_i}}{(2\pi)^D}
   \sum_{c=1}^{L(L+9)/2} \sum_{T\in P_c}\frac{\Delta_{c;T}(\{x_{ij}, \mu_{ij}\})}{\prod_{l\in T} D_l(\{k\},\{p\})} \;,
\end{align}
where $x_{ij}$ and $\mu_{ij}$ are 4 and $-2\eps$ dimensional irreducible scalar products (ISPs) that must be identified. To achieve this we use the polynomial division algorithm proposed by Zhang \cite{Zhang:2012ce} in $D$-dimensions:
\begin{enumerate}
  \item Choose a maximal propagator topology from the list of un-computed topologies.
  \item Choose a set of momenta, $v_i$, spanning the space of external momenta.
  \item Change the propagator equations into scalar product variables $x_{ij} = \bar{k}_i \cdot v_j$ and $\mu_{ij} = k_i^{[-2\eps]} \cdot k_j^{[-2\eps]}$
    using the $4\times4$ Gram matrix, $G_4(v)$, to give a set of equations: $P_\alpha(x_{ij}, \mu_{ij})$.
  \item Separate the reducible scalar products from the irreducible scalar products using the linear parts of the
    propagator equations, $\la P \ra = \la P_{\rm quadratic} \ra \cup \la P_{\rm linear} \ra$.
  \item Use polynomial division by the Gr\"obner basis of the ideal $\la P_{\rm quadratic} \ra$ to define
    the integrand parametrization $\Delta_{T} = \sum_i c_i m_i(x_{ij}, \mu_{ij})$ in terms of ISP monomials $m_i$.
  \item Use primary decomposition of the algebraic variety $Z(P)$ to reduce the
    unitarity-cut solutions to irreducible branches. Solve the
    on-shell equations at each branch using an explicit parametrization of $k^{(s)}_i(\tau_\alpha)$.
  \item Use $\Delta_{T}(k^{(s)}) = \sum d_j m'_j(\tau_\alpha) = N(k^{(s)}) - \sum_T \Delta_T(k^{(s)})  = \prod A^{(0)}(k^{(s)}) - \sum_T \Delta_T(k^{(s)})$ to compute
    the residues $d_j$ from the input (e.g. tree-level amplitudes, diagrams, etc). The sum runs over all (previously computed)
    topologies with higher number of propagators
  \item Solve the master system $M\cdot \vec{c} = \vec{d}$ for the coefficients $c_i$ where,
    \begin{align}
      \sum_j d_j m'_j(\tau_\alpha) = \sum_i c_i m_i(x_{ij}(\tau_\alpha), \mu_{ij}(\tau_\alpha)) = \sum_{i,j} M_{ji} c_i m'_j(\tau_\alpha)
      \label{eq:linearsysdef}
    \end{align}
  \item Go back to 1 until all topologies are computed.
\end{enumerate}

Steps 2-5 above are implemented in Zhang's {\sc BasisDet} Mathematica package \cite{Zhang:2012ce}.

There are a few important comments to make about this approach. Firstly there is
an ambiguity as to which spanning basis vectors $v_i$ to choose and we notice
in a number of examples that a good choice can result in a considerably more
compact integrand representation. Secondly, step 6 can be
efficiently performed using the primary decomposition algorithm implemented in
{\sc Macaulay2} \cite{M2}. In $D$-dimensions it is possible to
prove that all propagator ideals are \textit{radical} ideals and therefore
admit exactly one on-shell solution branch. Thirdly, in step 5 we must provide
an ISP ordering to define the polynomial division. Any choice will result in a valid
integrand representation but also here the choice will affect the analytic form considerably.
A feature of $D$-dimensional systems is that we may elect to remove monomials of the extra dimensional
ISPs $\mu_{ij}$ in terms of monomials of the four dimensional ones $x_{ij}$. In certain circumstances
this can result in a $D$-dimensional integrand basis which does not a have a smooth limit
onto the four dimensional case, and this is a feature which is highly undesirable and may require
some additional manipulation after the polynomial division has been performed.

\section{Simplifying kinematics with momentum twistors}

The four-component momentum twistors, first introduced by Hodges \cite{Hodges:2009hk}, is $Z(p_i) = Z_i = (\lambda_{i,a},
\mu_{i,a})$ for a massless momentum $p_i$, where $\lambda_{i,a}$ are the two-component holomorphic Weyl spinors ($a=1,2$). $\mu_i$
are the dual variables which are used to construct the anti-holomorphic spinors:
\begin{align}
  W_{i,\dot{a}} = (\tilde{\mu}_{i,\dot{a}}, \tilde{\lambda}_{i,\dot{a}}) =
  \frac{\epsilon_{\dot{a}, b, c, d} Z_{i-1,b} Z_{i,c} Z_{i+1,d}}{\langle i-1, i \rangle\langle i, i+1 \rangle}.
\end{align}
While for four-point kinematics the minimal set of variables can be written in terms of Mandelstam invariants, the
five-point case is complicated by a non-trivial Gram matrix identity relating the square of the trace
operator including $\gamma_5$ to the invariants,
\begin{align}
  \tr_5(1234)^2 = 16 \det G \!
\begin{pmatrix}
  p_1 & p_2 & p_3 & p_4 \\
  p_1 & p_2 & p_3 & p_4
\end{pmatrix}.
\end{align}
This relation is satisfied explicitly after a transformation to the momentum twistor variables, $x_i$:
\begin{align}
  Z =
\begin{pmatrix}
  1 & 0 & \tfrac{1}{x_1} & \tfrac{1}{x_1} + \tfrac{1}{x_1 x_2} & \tfrac{1}{x_1} + \tfrac{1}{x_1 x_2} + \tfrac{1}{x_1 x_2 x_3} \\
  0 & 1 & 1 & 1 & 1 \\
  0 & 0 & 0 & \tfrac{x_4}{x_2} & 1 \\
  0 & 0 & 1 & 1 & 1-\tfrac{x_5}{x_4}
\end{pmatrix}.
\end{align}
The variables $x_i$ can be expressed in terms of the usual spinor products and kinematic invariants as:
\begin{align}
  x_1 &= s_{12} \;, &
  x_4 &= \frac{s_{23}}{s_{12}} \;, &
  x_5 &= \frac{s_{123}}{s_{12}} \;, \nonumber \\
  x_2 &= -\frac{\spA23\spA41}{\spA12\spA34} \;, &
  x_3 &= -\frac{\spA34\spA51}{\spA13\spA45} \;.
\end{align}
By converting expressions to these variables during step 7 of the algorithm, we can automatically
express the Feynman diagram input as compact analytic expressions for the $d_j$ coefficients.

\section{Example: non-planar maximal cuts of five-gluon all-plus helicity amplitude}

\begin{figure}
\begin{subfigure}{.5\textwidth}
  \begin{center}
    \includegraphics[height=3cm]{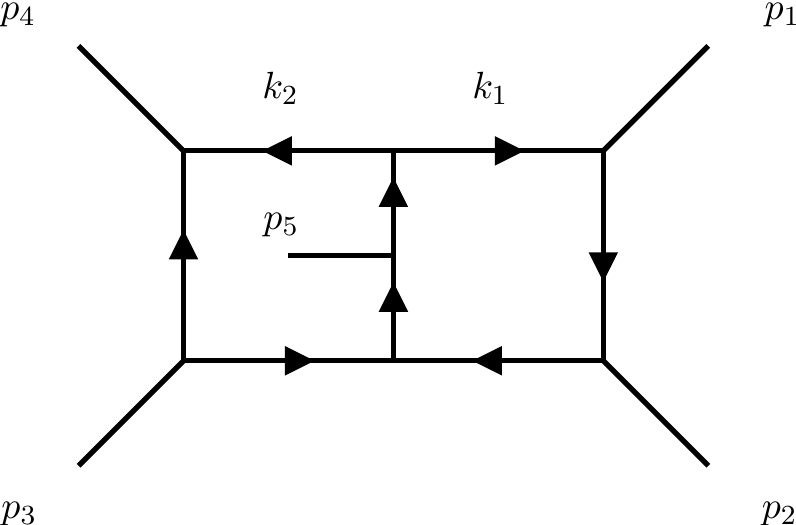}
  \end{center}
  \caption{The topology $(332)$.}
  \label{fig:db332}
\end{subfigure}
\begin{subfigure}{.5\textwidth}
  \begin{center}
    \includegraphics[height=3cm]{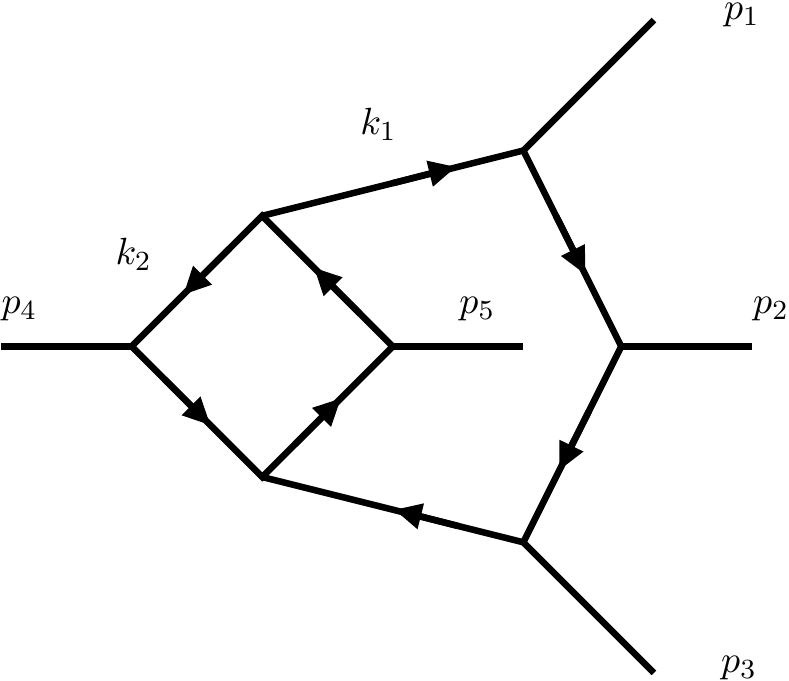}
  \end{center}
  \caption{The topology $(422)$.}
  \label{fig:db422}
\end{subfigure}
  \caption{Maximal cut non-planar topologies for five-point amplitudes.}
\label{fig:nptopos}
\end{figure}

There are two independent non-planar topologies for massless five-point amplitudes as depicted in
figure \ref{fig:nptopos}. We label each topology by the number of propagators along each of the
three loop momentum branches $k_1$, $k_2$ and $k_1+k_2$ so topology (a) is $332$ and (b) is $422$.

\subsection{The $332$ topology}

This topology is defined by the propagators:
\begin{equation}
  \{k_1, \;
  k_1-p_1, \;
  k_1-p_1-p_2, \;
  k_2, \;
  k_2-p_4, \;
  k_2-p_3-p_4, \;
  k_1+k_2, \;
  k_1+k_2+p_5\}.
  \label{eq:332props}
\end{equation}
A good choice for the spanning basis turns out to be $v=\{5,1,4,2\}$ which, using
a lexicographic ordering, gives a generic basis of 82 monomials under the renormalization
constraints $\{5,5,6\}$. The {\sc BasisDet} code is rather simple and reads:
\begin{Verbatim}[fontsize=\tiny]
L=2;
Dim=4-2\[Epsilon];
n=5;
ExternalMomentaBasis = {p5, p1, p4, p2};
Kinematics = {
   p1^2 -> 0, p2^2 -> 0, p4^2 -> 0, p5^2 -> 0,
   p1*p2 -> s12/2, p1*p3 -> (s45-s12-s23)/2, p1*p4 -> (s23-s15-s45)/2, p1*p5 -> s15/2,
   p2*p3 -> s23/2, p2*p4 -> (s15-s23-s34)/2, p2*p5 -> (s34-s12-s15)/2,
   p3*p4 -> s34/2, p3*p5 -> (s12-s34-s45)/2,
   p4*p5 -> s45/2
};
numeric = {s12 -> 11, s23 -> 17, s34 -> 7, s45 -> 3, s15 -> 29};
Props = {l1, l1 - p1, l1 - p1 - p2, l2, l2 - p4, l2 - p3 - p4, l1 + l2 + p5 , l1 + l2};
RenormalizationCondition = {{{1, 0}, 5}, {{0, 1}, 5}, {{1, 1}, 6}};
GenerateBasis[0]
\end{Verbatim}
The on-shell loop momenta are conveniently parametrized in terms of 3 free variables:
\begin{align}
  \bar{k}_1^\mu &= p_1^\mu + \tau_1 \frac{\spB13}{2\spB23} \cv12 + \tau_2 \frac{\spA13}{2\spA23} \cv21 \;, \\
  \bar{k}_2^\mu &= p_4^\mu + \tau_3 \frac{\spA14}{2\spA13} \cv34 + \beta \frac{\spA13}{2\spA14} \cv43 \;,
  \label{eq:db332-osmom}
\end{align}
where
\begin{equation}
  \beta = \frac{-s_{14}}{\trm(1354)}\left(
    s_{15} + s_{45}
    + \frac{1}{s_{23}} \big( \tau_1\trm(1523)+\tau_2\trp(1523) \big)
    + \frac{1}{s_{13}}\tau_3 \trm(1453)\right).
\end{equation}
We can then construct a linear system of $82\times83$ to fit the integrand coefficients. In the case of the
5-gluon all-plus amplitude the situation is much simpler than this general parametrization suggests, and we find that only
three independent ISP monomials are present with a structure similar to that of the planar case \cite{Badger:2013gxa}. The result is:
\begin{align}
  \Delta_{8;332}(1^+,2^+,3^+,4^+,5^+) &= \frac{ 2 iF_1 \; s_{12}s_{34}}{\spA12\spA23\spA34\spA45\spA51 \tr_5}
  \left( c_1 \MP{k_1}{p_4} + c_2 \MP{k_2}{p_1} + c_3 \MP{k_1}{p_5}
  \right), \\
  c_1 &= -s_{15}\trm(2345), \\
  c_2 &= s_{45}\trm(2351), \\
  c_3 &= s_{23} s_{45} s_{15} - s_{15}\trm(2345) - s_{45}\trm(2351), \\
  F_1 &= (D_s-2)(\mu_{11}\mu_{22}+\mu_{22}\mu_{33}+\mu_{33}\mu_{11}) + 4(\mu_{12}^2-4\mu_{11}\mu_{22}),
  \label{eq:Delta332}
\end{align}
where $\mu_{33} = \mu_{11}+\mu_{12}+\mu_{22}$, and $D_s = g^\mu_{\;\, \mu}$ is the spin dimension.

\subsection{The $422$ topology}

This topology is defined by the propagators:
\begin{equation}
  \{k_1, \;
  k_1-p_1, \;
  k_1-p_1-p_2, \;
  k_1-p_1-p_2-p_3, \;
  k_2, \;
  k_2-p_4, \;
  k_1+k_2, \;
  k_1+k_2+p_5\},
  \label{eq:422props}
\end{equation}
and again in this case we use $v=\{5,1,4,2\}$. To make the four-dimensional limit of the integrand
representation manifest we must make some replacements to the monomial list. One possible change is:
\begin{align}
  x_{22}^4 &\to x_{22}^4 \mu_{22}, &
  x_{22}^3 &\to x_{22}^3 \mu_{22}, &
  x_{22}^2 &\to x_{22}^2 \mu_{11}, \nonumber \\
  x_{24} x_{22}^3 &\to x_{24} x_{22}^3 \mu_{22}, &
  x_{24} x_{22}^2 &\to x_{24} x_{22}^2 \mu_{22}, &
  x_{24} x_{22} &\to x_{24} x_{22} \mu_{11}, \nonumber \\
  x_{21} x_{22} &\to x_{21} x_{22} \mu_{22}^2.
\end{align}
The on-shell solution can be parametrized as:
\begin{align}
  \bar{k}_1^\mu &= p_1^\mu + \tau_1 \frac{\spA23}{2\spA13} \cv12 + (1-\tau_1) \frac{\spB23}{2\spB13} \cv21 \;, \\
  \bar{k}_2^\mu &= \beta p_4^\mu + \tau_2 \frac{\spA15}{2\spA14} \cv45 + \tau_3 \frac{\spB15}{2\spB14} \cv54 \;,
  \label{eq:db422-osmom}
\end{align}
where
\begin{align}
  \beta = -\frac{1}{s_{45}s_{13}}\left( s_{13}s_{15} + \tau_1\trm(1523) + (1-\tau_1)\trp(1523) \right).
\end{align}
Inverting the resulting $65\times76$ linear system gives a simple representation for the all-plus helicity configuration:
\begin{align}
  \Delta_{8;422} &= \frac{ iF_1 \; s_{12}s_{23}s_{45}}{\spA12\spA23\spA34\spA45\spA51 \tr_5}
  \left( c_0 + 2 c_1 \MP{k_2}{p_5} \right), \\
  c_0 &= s_{15}s_{34}s_{45},\\
  c_1 &= -\trp(1345).
  \label{eq:Delta422}
\end{align}
where $F_1$ refers to eq. (\ref{eq:Delta332}).

As observed in the case of the planar topologies \cite{Badger:2013gxa}, these non-planar contributions also have compact representations for the all-plus
configuration. Though more general QCD helicity amplitudes will be significantly more complicated, we hope that the techniques
presented here will help make these computations possible in the near future.


\begin{thebibliography}{10}

\bibitem{Ridder:2013mf}
A.~G.-D. Ridder, T.~Gehrmann, E.~Glover, and J.~Pires {\em Phys.Rev.Lett.} {\bf
  110} (2013) 162003, [\href{http://xxx.lanl.gov/abs/1301.7310}{{\tt
  1301.7310}}].

\bibitem{Currie:2013dwa}
J.~Currie, A.~Gehrmann-De~Ridder, E.~Glover, and J.~Pires {\em JHEP} {\bf 1401}
  (2014) 110, [\href{http://xxx.lanl.gov/abs/1310.3993}{{\tt 1310.3993}}].

\bibitem{Abelof:2014fza}
{Abelof, Gabriel and Gehrmann-De Ridder, Aude and Maierh\"{o}fer, Philipp and
  Pozzorini, Stefano} \href{http://xxx.lanl.gov/abs/1404.6493}{{\tt
  1404.6493}}.

\bibitem{Czakon:2013goa}
M.~Czakon, P.~Fiedler, and A.~Mitov {\em Phys.Rev.Lett.} {\bf 110} (2013)
  252004, [\href{http://xxx.lanl.gov/abs/1303.6254}{{\tt 1303.6254}}].

\bibitem{Cascioli:2014yka}
{Cascioli, F. and Gehrmann, T. and Grazzini, M. and Kallweit, S. and
  Maierh\"{o}fer, P. and others} \href{http://xxx.lanl.gov/abs/1405.2219}{{\tt
  1405.2219}}.

\bibitem{Chetyrkin:1981qh}
K.~Chetyrkin and F.~Tkachov {\em Nucl.Phys.} {\bf B192} (1981) 159--204.

\bibitem{Anastasiou:2000kg}
C.~Anastasiou, E.~Glover, C.~Oleari, and M.~Tejeda-Yeomans {\em Nucl.Phys.}
  {\bf B601} (2001) 318--340,
  [\href{http://xxx.lanl.gov/abs/hep-ph/0010212}{{\tt hep-ph/0010212}}].

\bibitem{Anastasiou:2000ue}
C.~Anastasiou, E.~Glover, C.~Oleari, and M.~Tejeda-Yeomans {\em Nucl.Phys.}
  {\bf B601} (2001) 341--360,
  [\href{http://xxx.lanl.gov/abs/hep-ph/0011094}{{\tt hep-ph/0011094}}].

\bibitem{Anastasiou:2001sv}
C.~Anastasiou, E.~Glover, C.~Oleari, and M.~Tejeda-Yeomans {\em Nucl.Phys.}
  {\bf B605} (2001) 486--516,
  [\href{http://xxx.lanl.gov/abs/hep-ph/0101304}{{\tt hep-ph/0101304}}].

\bibitem{Glover:2001af}
E.~Glover, C.~Oleari, and M.~Tejeda-Yeomans {\em Nucl.Phys.} {\bf B605} (2001)
  467--485, [\href{http://xxx.lanl.gov/abs/hep-ph/0102201}{{\tt
  hep-ph/0102201}}].

\bibitem{Gehrmann:2011aa}
T.~Gehrmann, M.~Jaquier, E.~Glover, and A.~Koukoutsakis {\em JHEP} {\bf 1202}
  (2012) 056, [\href{http://xxx.lanl.gov/abs/1112.3554}{{\tt 1112.3554}}].

\bibitem{Bern:1994zx}
Z.~Bern, L.~J. Dixon, D.~C. Dunbar, and D.~A. Kosower {\em Nucl.Phys.} {\bf
  B425} (1994) 217--260, [\href{http://xxx.lanl.gov/abs/hep-ph/9403226}{{\tt
  hep-ph/9403226}}].

\bibitem{Britto:2004nc}
R.~Britto, F.~Cachazo, and B.~Feng {\em Nucl.Phys.} {\bf B725} (2005) 275--305,
  [\href{http://xxx.lanl.gov/abs/hep-th/0412103}{{\tt hep-th/0412103}}].

\bibitem{Bern:2000dn}
Z.~Bern, L.~J. Dixon, and D.~Kosower {\em JHEP} {\bf 0001} (2000) 027,
  [\href{http://xxx.lanl.gov/abs/hep-ph/0001001}{{\tt hep-ph/0001001}}].

\bibitem{Bern:2000ie}
Z.~Bern, L.~J. Dixon, and A.~Ghinculov {\em Phys.Rev.} {\bf D63} (2001) 053007,
  [\href{http://xxx.lanl.gov/abs/hep-ph/0010075}{{\tt hep-ph/0010075}}].

\bibitem{Bern:2001df}
Z.~Bern, A.~De~Freitas, and L.~J. Dixon {\em JHEP} {\bf 0109} (2001) 037,
  [\href{http://xxx.lanl.gov/abs/hep-ph/0109078}{{\tt hep-ph/0109078}}].

\bibitem{Bern:2001dg}
Z.~Bern, A.~De~Freitas, L.~J. Dixon, A.~Ghinculov, and H.~Wong {\em JHEP} {\bf
  0111} (2001) 031, [\href{http://xxx.lanl.gov/abs/hep-ph/0109079}{{\tt
  hep-ph/0109079}}].

\bibitem{Bern:2002tk}
Z.~Bern, A.~De~Freitas, and L.~J. Dixon {\em JHEP} {\bf 0203} (2002) 018,
  [\href{http://xxx.lanl.gov/abs/hep-ph/0201161}{{\tt hep-ph/0201161}}].

\bibitem{Bern:2002zk}
Z.~Bern, A.~De~Freitas, L.~J. Dixon, and H.~Wong {\em Phys.Rev.} {\bf D66}
  (2002) 085002, [\href{http://xxx.lanl.gov/abs/hep-ph/0202271}{{\tt
  hep-ph/0202271}}].

\bibitem{Bern:2003ck}
Z.~Bern, A.~De~Freitas, and L.~J. Dixon {\em JHEP} {\bf 0306} (2003) 028,
  [\href{http://xxx.lanl.gov/abs/hep-ph/0304168}{{\tt hep-ph/0304168}}].

\bibitem{Kosower:2011ty}
D.~A. Kosower and K.~J. Larsen {\em Phys.Rev.} {\bf D85} (2012) 045017,
  [\href{http://xxx.lanl.gov/abs/1108.1180}{{\tt 1108.1180}}].

\bibitem{Johansson:2012zv}
H.~Johansson, D.~A. Kosower, and K.~J. Larsen {\em Phys.Rev.} {\bf D87} (2013)
  025030, [\href{http://xxx.lanl.gov/abs/1208.1754}{{\tt 1208.1754}}].

\bibitem{Johansson:2013sda}
H.~Johansson, D.~A. Kosower, and K.~J. Larsen
  \href{http://xxx.lanl.gov/abs/1308.4632}{{\tt 1308.4632}}.

\bibitem{CaronHuot:2012ab}
S.~Caron-Huot and K.~J. Larsen {\em JHEP} {\bf 1210} (2012) 026,
  [\href{http://xxx.lanl.gov/abs/1205.0801}{{\tt 1205.0801}}].

\bibitem{Larsen:2012sx}
K.~J. Larsen {\em Phys.Rev.} {\bf D86} (2012) 085032,
  [\href{http://xxx.lanl.gov/abs/1205.0297}{{\tt 1205.0297}}].

\bibitem{Sogaard:2013yga}
{S\o gaard, Mads} {\em JHEP} {\bf 1309} (2013) 116,
  [\href{http://xxx.lanl.gov/abs/1306.1496}{{\tt 1306.1496}}].

\bibitem{Sogaard:2013fpa}
{S\o gaard, Mads and Zhang, Yang} {\em JHEP} {\bf 1312} (2013) 008,
  [\href{http://xxx.lanl.gov/abs/1310.6006}{{\tt 1310.6006}}].

\bibitem{Sogaard:2014ila}
{S\o gaard, Mads and Zhang, Yang} \href{http://xxx.lanl.gov/abs/1403.2463}{{\tt
  1403.2463}}.

\bibitem{Sogaard:2014oka}
{S\o gaard, Mads and Zhang, Yang} \href{http://xxx.lanl.gov/abs/1406.5044}{{\tt
  1406.5044}}.

\bibitem{Ossola:2006us}
G.~Ossola, C.~G. Papadopoulos, and R.~Pittau {\em Nucl.Phys.} {\bf B763} (2007)
  147--169, [\href{http://xxx.lanl.gov/abs/hep-ph/0609007}{{\tt
  hep-ph/0609007}}].

\bibitem{Mastrolia:2011pr}
P.~Mastrolia and G.~Ossola {\em JHEP} {\bf 1111} (2011) 014,
  [\href{http://xxx.lanl.gov/abs/1107.6041}{{\tt 1107.6041}}].

\bibitem{Badger:2012dp}
S.~Badger, H.~Frellesvig, and Y.~Zhang {\em JHEP} {\bf 1204} (2012) 055,
  [\href{http://xxx.lanl.gov/abs/1202.2019}{{\tt 1202.2019}}].

\bibitem{Zhang:2012ce}
Y.~Zhang {\em JHEP} {\bf 1209} (2012) 042,
  [\href{http://xxx.lanl.gov/abs/1205.5707}{{\tt 1205.5707}}].

\bibitem{Mastrolia:2012an}
P.~Mastrolia, E.~Mirabella, G.~Ossola, and T.~Peraro {\em Phys.Lett.} {\bf
  B718} (2012) 173--177, [\href{http://xxx.lanl.gov/abs/1205.7087}{{\tt
  1205.7087}}].

\bibitem{Badger:2012dv}
S.~Badger, H.~Frellesvig, and Y.~Zhang {\em JHEP} {\bf 1208} (2012) 065,
  [\href{http://xxx.lanl.gov/abs/1207.2976}{{\tt 1207.2976}}].

\bibitem{Mastrolia:2012wf}
P.~Mastrolia, E.~Mirabella, G.~Ossola, and T.~Peraro {\em Phys.Rev.} {\bf D87}
  (2013) 085026, [\href{http://xxx.lanl.gov/abs/1209.4319}{{\tt 1209.4319}}].

\bibitem{Kleiss:2012yv}
R.~H. Kleiss, I.~Malamos, C.~G. Papadopoulos, and R.~Verheyen {\em JHEP} {\bf
  1212} (2012) 038, [\href{http://xxx.lanl.gov/abs/1206.4180}{{\tt
  1206.4180}}].

\bibitem{Feng:2012bm}
B.~Feng and R.~Huang {\em JHEP} {\bf 1302} (2013) 117,
  [\href{http://xxx.lanl.gov/abs/1209.3747}{{\tt 1209.3747}}].

\bibitem{Mastrolia:2013kca}
P.~Mastrolia, E.~Mirabella, G.~Ossola, and T.~Peraro {\em Phys.Lett.} {\bf
  B727} (2013) 532--535, [\href{http://xxx.lanl.gov/abs/1307.5832}{{\tt
  1307.5832}}].

\bibitem{Huang:2013kh}
R.~Huang and Y.~Zhang {\em JHEP} {\bf 1304} (2013) 080,
  [\href{http://xxx.lanl.gov/abs/1302.1023}{{\tt 1302.1023}}].

\bibitem{Badger:2013gxa}
S.~Badger, H.~Frellesvig, and Y.~Zhang {\em JHEP} {\bf 1312} (2013) 045,
  [\href{http://xxx.lanl.gov/abs/1310.1051}{{\tt 1310.1051}}].

\bibitem{Feng:2014nwa}
B.~Feng, J.~Zhen, R.~Huang, and K.~Zhou
  \href{http://xxx.lanl.gov/abs/1401.6766}{{\tt 1401.6766}}.

\bibitem{Badger:2013sta}
S.~Badger, H.~Frellesvig, and Y.~Zhang {\em J.Phys.Conf.Ser.} {\bf 523} (2014)
  012061, [\href{http://xxx.lanl.gov/abs/1310.4445}{{\tt 1310.4445}}].

\bibitem{M2}
D.~R. Grayson and M.~E. Stillman, {\it \textsc{Macaulay2}, a software system
  for research in algebraic geometry}. Available at
  \href{http://www.math.uiuc.edu/Macaulay2/}{http://www.math.uiuc.edu/Macaulay2/}.

\bibitem{Hodges:2009hk}
A.~Hodges {\em JHEP} {\bf 1305} (2013) 135,
  [\href{http://xxx.lanl.gov/abs/0905.1473}{{\tt 0905.1473}}].

\end{thebibliography}
\providecommand{\href}[2]{#2}\begingroup\raggedright\endgroup
\end{document}